\documentclass{aa}
\usepackage{graphicx}
\usepackage{txfonts}
\usepackage{hyperref}
\hypersetup{colorlinks=true, urlcolor=blue, citecolor=cyan, pdfborder={0 0 0}}

\begin{document}

\title{pyUPMASK: An improved unsupervised clustering algorithm}

\author{M. S. Pera\inst{1}
\and
G. I. Perren\inst{1}
\and
A. Moitinho\inst{2}
\and
H. D. Navone\inst{3,4}
\and
R. A. Vazquez\inst{5}
}

\institute{Instituto de Astrof\'isica de La Plata (IALP-CONICET), 1900 La
Plata, Argentina\\
\email{msolpera@gmail.com}
\and
CENTRA, Faculdade de Ci\^encias, Universidade de Lisboa, Ed. C8, Campo Grande,
1749-016 Lisboa, Portugal
\and
Facultad de Ciencias Exactas, Ingenier\'ia y Agrimensura (UNR), 2000 Rosario,
Argentina
\and
Instituto de F\'isica de Rosario (CONICET-UNR), 2000 Rosario, Argentina,
\and
Facultad de Ciencias Astronómicas y Geofísicas (UNLP-IALP-CONICET), 1900 La
Plata, Argentina
}

\date{Received December 28, 2020; accepted March 25, 2021}

\abstract
  {}
  {We present pyUPMASK, an unsupervised clustering method for stellar clusters
  that builds upon the original UPMASK package. The general approach of this method makes it
  plausible to be applied to analyses that deal with binary classes of any kind as long as the fundamental hypotheses are met.
  The code is written entirely in Python and is made available through a public
  repository.}
  {The core of the algorithm follows the method developed in UPMASK but introduces several key enhancements. These enhancements not only make
  pyUPMASK more general, they also improve its performance considerably.
  }
  {We thoroughly tested the performance of pyUPMASK on 600 synthetic clusters affected by varying degrees of contamination by field stars. To assess the
  performance, we employed six different statistical metrics that measure the
  accuracy of probabilistic classification.}
  {Our results show that pyUPMASK is better performant than UPMASK for every
  statistical performance metric, while still managing to be many times faster.
  }


\maketitle

\section{Introduction}
 \label{sec:intro}

 Galactic open clusters are of great importance for the study of the
 Galaxy's chemical evolution, structure, and dynamics; these sources also provide test beds for astrophysical codes that model the evolution of stars.
 Located largely on the disk of the Milky Way, analyses of open clusters is severely
 hindered by the presence of contaminating field stars, located in the
 foreground and background of the object of interest. These stars are
 projected on the observed field of view and end up deeply mixed
 with the cluster members. The process of disentangling these two classes of
 elements, of members from nonmembers (i.e., field stars), can be referred
 to as ``decontamination''.
 A proper decontamination of the cluster region is a key previous step to the
 analysis of the cluster sequence in search of fundamental parameters 
 (e.g., metallicity, age, distance and extinction) that characterize the open
 cluster. This analysis, which is often performed in photometric
 space, requires a sequence that is as complete as possible, but also as free
 of contaminating field stars (nonmembers) as possible. The goal of a
 decontamination algorithm is to obtain a subset of stars that fulfills both
 these conditions simultaneously.

 Over the years, a handful of decontamination algorithms have been presented in the
 stellar cluster literature. Most of these are variations of the
 Vasilevskis-Sanders method \citep{Vasilevskis_1958,Sanders_1971} applied over
 proper motions, which are generally considered to be much better member
 discriminators than photometry.
 Nonparametric approaches have also been developed
 \citep{Cabrera_1990,Javakhishvili_2006} and even an interactive tool to
 determine membership probabilities was presented
 \citep{Balaguer_2020}\footnote{Clusterix 2.0:
 \url{http://clusterix.cab.inta-csic.es/clusterix/}}.
 More references regarding membership estimation methods can be found in
 \citet[][henceforth KMM14]{KMM14} and \citet{Perren_2015}.

 The Unsupervised Photometric Membership Assignment in Stellar Clusters
 algorithm (UPMASK), originally presented in KMM14,  has the advantage of being
 not only nonparametric, but also unsupervised.  This means that no a priori
 selection of field stars is required to serve as a  comparison model, which is
 generally the case in the previously mentioned methods. Although UPMASK was motivated by the
 need of assigning cluster memberships from photometric data, KMM14 had pointed
 out that the method is general and could be easily applied to other data
 types and clusters of objects. Recent examples of UPMASK used
 on proper motions (and parallax data) can be found in \citet{Cantat_2018},
 \citet{Cantat_2018_2}, \citet{Cantat_2019}, \citet{Carrera_2019}, and
 \citet{Yontan_2019}.
 In the six years since its publication, the KMM14 article has been referenced almost 50 times;
 this work has also been applied to stellar proper motions and to study clusters of galaxies, 
 which indicates a wide adoption by the astrophysical community.

 In this work we present an improved version of the original UPMASK algorithm, which we
 call pyUPMASK because it is written entirely in Python. We believe this new package
 can be of great use, particularly with the advent of the recent early data
 release 3 \citep[eDR3,][]{GaiaEDR3_2020} of the Gaia
 mission \citep{Gaia_2016}.
 This package is made available as a stand-alone code, but it will also be
 included in an upcoming release of our Automated Stellar Cluster Analysis
 tool \citep[\texttt{ASteCA},][]{Perren_2015}.
 Throughout the article we refer to statistical clusters as simply
 clusters and explicitly distinguish them from stellar clusters when
 required.\\

 This paper is organized as follows: In Section \ref{sec:methods} we
 give a brief summary of the UPMASK algorithm and present the details of the
 enhancements introduced in our code. 
 Section \ref{sec:validation} introduces the synthetic cluster sample
 used in the analysis, and describes the selected statistical performance metrics
 employed to assess the behavior of UPMASK and pyUPMASK.
 The results are summarized in Section \ref{sec:results}. Finally, our
 conclusions are given in Section \ref{sec:conclusions}.

\section{Methods}
 \label{sec:methods}

 We present a brief description of the general algorithm used in UPMASK
 as well as the major enhancements introduced in pyUPMASK.
 Both methods are open source and their codes can be found in their
 respective public repositories.\footnote{UPMASK: 
 \url{https://cran.r-project.org/web/packages/UPMASK/}}$^{,}$\footnote{pyUPMASK:
 \url{https://github.com/msolpera/pyUPMASK}}

\subsection{The UPMASK algorithm}
 \label{ssec:upmask}

 The UPMASK package is described in full in KMM14 and we do not repeat it in
 this work. We give instead a summary of the most relevant parts and of its core
 algorithm. The original article provides a more detailed
 description.\\
 
 Assigning probability memberships to the two classes of elements within a 
 stellar cluster field (members and field stars) is a notably complicated
 problem for two main reasons. First, the classes are usually very much
 imbalanced. This means that one of the classes (field stars) can make up a
 lot more than 50\% of the total dataset. In some extreme cases, the frame of
 an observed stellar cluster can consist of over 90\% of field stars and less
 than 10\% of actual true members. Even worse, this information (i.e., the
 true balance) cannot be assumed to be known a priori.
 Second, the two classes are deeply entangled. This is particularly true in the
 two-dimensional coordinates space where members and field stars are mixed
 throughout the entire cluster region. Off the shelf clustering methods
 normally assume that there is some kind of frontier that largely separates
 the classes with minimal overlap. This is not the case in stellar clusters
 analysis.
 The UPMASK algorithm deals with both of these issues in a clever and effective way, by taking
 advantage of the fact that we can approximate the distribution of field stars
 in the coordinates space with a uniform model. This is further discussed
 in Sect.~\ref{sssec:ripley}.\\

 The UPMASK algorithm is composed of two main blocks: an outer loop and an inner
 loop. The outer loop is responsible for taking into account the uncertainties in the data and rerunning the inner loop a manually fixed number of times; these uncertainties are optional and turned off by default. The latter is required because
 of the inherent stochasticity of the K-means (KMS) method \citep{macqueen1967},
 employed by the inner loop. The number of runs for the outer loop is one of
 the two most important parameters in the algorithm.
 The inner loop holds the two main parts that make up the core of the
 algorithm: the clustering method (KMS, as stated before), and the
 random field rejection method (henceforth: RFR).
 The clustering method is applied on the nonpositional features (e.g. photometry and
 proper motions), and separates the cluster data into $N$ clusters. The
 $N$ value is determined by a parameter
 that determines the number of elements that should be
 contained in each cluster. That is, dividing the total number of stars by this
 value gives $N$, the final number of clusters that are generated.
 
 After the clustering method is applied, the RFR method serves the purpose of
 filtering those clusters identified by the KMS that are consistent
 with a random uniform distribution of elements. This consistency is assessed
 in UPMASK by means of a two-dimensional kernel density estimation (KDE)
 analysis. In short: the KDE of the coordinates space of each cluster 
 (identified by the KMS in the previous step) is compared with the KDE of a
 two-dimensional uniform distribution in the same range. If these are deemed
 to be similar enough, the cluster is discarded as a realization of a random
 selection of field stars, and all its stars are assigned a value of 0. Those
 clusters that survive the RFR process are kept for a subsequent iteration of
 the inner loop. When no more clusters are rejected and the inner loop is finished,
 all the stars within surviving clusters are assigned a value of 1. After this,
 a new iteration of the outer loop is initiated.
 The final probabilities assigned to each star are simply the averages of the
 (0, 1) values assigned by the inner loop at each run of the outer loop.\\

 The two parameters mentioned above are the most important parameters in
 UPMASK, since varying their value can substantially affect the performance
 of the method. We comment on how we selected these
 parameters in Sect~\ref{ssec:input_pars}.

\subsection{The pyUPMASK algorithm}
 \label{ssec:pyupmask}

  An obvious difference between pyUPMASK and UPMASK is that the former is
  written entirely in Python\footnote{\url{https://www.python.org/}} instead
  of R\footnote{\url{https://www.r-project.org/}}, as is the case with UPMASK.
  We believe that this is a considerable advantage given the noticeable shift
  of the astrophysical community toward the Python language in recent years.
  This is made evident by large Python-based projects such as
  Astropy\footnote{\url{http://www.astropy.org}} \citep{astropy:2013,
  astropy:2018}
  and international conferences such as Python in
  Astronomy\footnote{\url{http://openastronomy.org/pyastro/}}. A recent
  survey found that Python is the most popular programming language in the
  astronomical community \citep{Momcheva2015,Tollerud2019SustainingCS}.

  \begin{figure}
   \centering
   \includegraphics[width=\hsize]{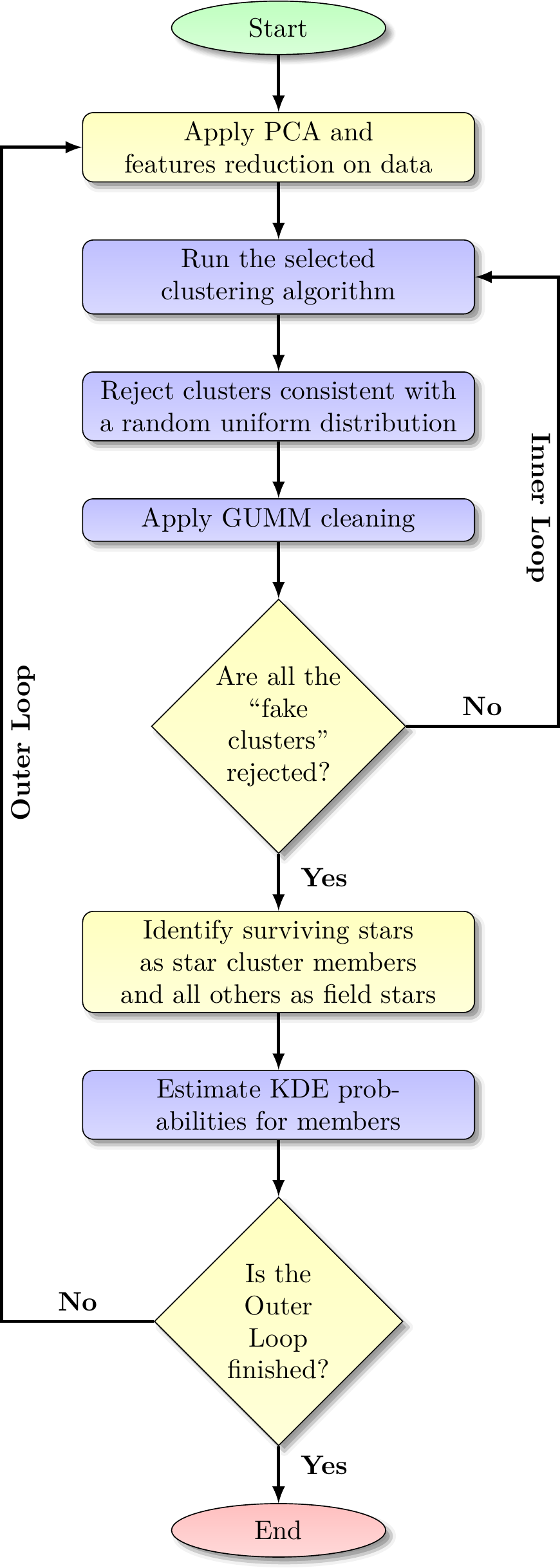}
   \caption{Flow chart of the pyUPMASK code. The enhanced clustering block and
   the analysis blocks added in this work are indicated in violet.}
   \label{fig:flowchart}
  \end{figure}

  The general structure of pyUPMASK closely follows the UPMASK algorithm: an outer 
  loop containing an inner loop that applies the cluster
  identification and rejection methods. What sets these two algorithms apart is twofold:
  First, pyUPMASK supports almost a dozen clustering
  methods, while UPMASK only supports KMS; and second, pyUPMASK contains
  three added analysis blocks that are not present in UPMASK.
  In Fig.~\ref{fig:flowchart} we show the complete flow chart of the pyUPMASK
  algorithm. The blocks indicated in violet are those that are either enhanced or
  added in this work.
  The enhanced clustering methods block and the three added blocks are detailed
  in Sects~\ref{sssec:clustering}, \ref{sssec:ripley},
  \ref{sssec:gumm}, and \ref{sssec:kde-probs}, respectively. The remaining
  portions of the code are mostly equivalent to those described in KMM14 for
  UPMASK and, for the sake of brevity, we do not repeat their details or
  purpose in this work.

\subsubsection{Clustering methods}
 \label{sssec:clustering}

 While UPMASK supports the KMS method exclusively (as of the current
 version 1.2), pyUPMASK relies on the Python library
 \texttt{scikit-learn}\footnote{\url{https://scikit-learn.org/}}
 \citep{scikit-learn} for the implementation of most of the supported
 clusterings methods. This library includes around a dozen different
 clustering methods for unlabeled data, which are all available to use in
 pyUPMASK. Eventually this can be extended to support even more methods
 in future releases of the code via the \texttt{PyClustering}
 library\footnote{\url{https://pyclustering.github.io}}.\\

 Once chosen, the clustering method processes the nonspatial data at the
 beginning of the inner loop as shown in Fig.~\ref{fig:flowchart}. The number
 of individual clusters to generate is fixed indirectly through a
 user-selected input parameter, as done in UPMASK.
 Each of these clusters is then analyzed by the RFR method and kept or
 rejected given its similarity with a random uniform distribution of elements.
 This is further discussed in Sect.~\ref{sssec:ripley}.\\

 In Sect~\ref{sec:results} we present a suit of tests performed with
 four of the methods provided by \texttt{scikit-learn}: KMS, mini
 batch k-means~\citep[MBK,][]{Sculley2010}, gaussian mixture models~
 \citep[GMM,][]{Baxter2010}, and agglomerative
 clustering~\citep[AGG,][]{Zepeda2013}.
 In addition to these we include tests performed with two methods developed in
 this work: the nearest neighbors density method (KNN), which is based on the
 density peak approach introduced in \cite{Rodriguez2014}; and the 
 Voronoi (VOR) method, which is based on the construction of N-dimensional
 Voronoi diagrams~\citep{Voronoi_1908}.
 The latter three methods (AGG, KNN and VOR) have a characteristic in common:
 no stochastic process or approximation is employed by any of them. In
 other words, these methods are deterministic.
 This means that, for the same input data and input parameters,
 different runs lead to one single result.
 Assuming that no data resampling is performed (the default setting in both
 UPMASK and pyUPMASK) the outer loop then needs to be run only once because
 subsequent runs would produce the same probabilities each time. For this
 reason we refer to these as ``single-run'' methods. As can easily be
 inferred, these are significantly faster than UPMASK and the rest of the
 tested methods, which require multiple outer loop runs.

 The results obtained with the six selected methods are compared to UPMASK
 results obtained on the same dataset of synthetic clusters. The synthetic
 clusters dataset is described in  Sect.~\ref{ssec:synthetic}.

\subsubsection{Ripley's K function}
 \label{sssec:ripley}

 After the clusters are generated on the nonspatial data, the RFR block is
 used to filter out those that are consistent with the realization of a random
 uniform distribution on the spatial data (i.e., coordinates). The hypothesis
 at work is that field stars are randomly scattered throughout the
 two spatial dimensions of the frame, following somewhat closely a uniform
 distribution. Actual star cluster members, on the other hand, present a
 more densely packed spatial distribution.
 The latter is of course an approximation to the real, and unknown,
 probability distribution of field stars, but it is still a very reasonable
 one, as the results show.

 The UPMASK algorithm employs a KDE-based method to characterize the distribution of each
 cluster found in the spatial dimensions. This distribution is then compared
 to that of thousands of random uniform distributions generated in the same
 two-dimensional range and with the same number of elements. After that,
 a ``KDE distance'' is obtained by comparing their means, maximum, and
 standard deviation values.
 If the distance between both distributions is less than a user-defined
 threshold parameter, the cluster is considered to be close enough to a
 realization of a random uniform distribution. When this condition is met, the
 cluster is rejected as a ``fake cluster'' (see Fig~\ref{fig:flowchart}).

 In pyUPMASK we introduce Ripley's K function \citep{ripley_1976,ripley_1979}
 to assess the closeness of a cluster to a random uniform distribution. This
 function is defined as

 \begin{equation}
 \hat{R}(r) = \frac{A}{N^2} \sum_i^N \sum_{j\neq i}^N I(d_{ij} < r) e_{ij},
 \end{equation}

 \noindent where $A$ is the area of the domain (our observed frame), $N$ is
 the number of points within it, $d_{ij}$ is the distance between points
 $i,j$, $I$ is a function that results in 1 if the condition is met and 0
 otherwise, $e_{ij}$ is the edge correction (if required), and $r$ is the
 scale at which the $ \hat{R}$ function is calculated.

 Ripley's K function is employed to test for complete spatial randomness (CSR),
 also called homogeneous Poisson point process, which basically consist of
 points randomly located on a given domain. In a two-dimensional space it is
 trivial to prove that if points are distributed following CSR, then $K(r)$
 equals $\pi r^2$~\citep{Streib_2011}.
 The K function is thus a perfect match for our intended usage which is
 precisely to test if a set of points (stars) are distributed following
 uniform spatial randomness.
 We employ the form of the K function given by

 \begin{equation}
 \hat{L}(r) = [\hat{K}(r)/\pi]^2,
 \end{equation}

 \noindent which converges to $r$ under CSR. Following~\cite{Dixon_2014} we
 combine information from several distances ($r$ values) in a single test
 statistic defined as

 \begin{equation}
 \hat{L}_{m} = \sup_{r} |\hat{L}(r) - r|,
 \label{eq:ripley_L}
 \end{equation}

 \noindent where sup is the supremum.
 Given that the lengths of the observed frame are normalized by default to the
 range $[0,1]$ prior to processing, the list of distances at which Eq.
 \ref{eq:ripley_L} is calculated are chosen to be in the range $[0, 0.25]$.
 This is the range advised in the \texttt{Kest} function of the
 \texttt{spatstat}
 package~\citep{Baddeley_2015}\footnote{\url{http://spatstat.org/}}.

 The null hypothesis ($H_0$) for the $\hat{L}_{m}$ is that the points follow
 CSR. We need to select a critical value such that if the test is
 greater than that value, the test is considered to be statistically
 significant and $H_0$ is rejected.
 Such critical values were estimated by Monte Carlo simulations
 in~\cite{ripley_1979}. The pyUPMASK algorithm uses the 1\% critical value; that is,
 there is a 1\% probability of erroneously rejecting $H_0$ (also called a
 Type I error). This critical value is approximated for $\hat{L}_{m}$ as $1.68\sqrt{A}/N$,
 where
 $A$ and $N$ are the area and number of points, respectively.
 In future releases of the code we plan on integrating analytical
 expressions for the critical values, for example, those obtained in
 \cite{Lagache_2013} and~\cite{Marcon_2013}.\\

 The pyUPMASK algorithm employs the \texttt{astropy} implementation of the K function, which
 includes the required edge corrections for points that are located close to
 the domain boundaries. Compared to the UPMASK KDE test, the K function is not
 only a more natural choice for this task, it is also orders of magnitude
 faster.

\subsubsection{Gaussian-Uniform mixture model}
 \label{sssec:gumm}

 After the RFR block is finished and the fake clusters are rejected, only
 those stars that were found in clusters sufficiently different from a random
 uniform distribution of points are kept. This dataset of stars is
 nonetheless still affected by contamination from field stars that could not be
 removed. This is because these field stars were, by chance, associated with a
 cluster composed mainly of true star cluster members and thus not rejected.
 We developed a method to clean this region, applied to the two-dimensional
 coordinates space that we call GUMM, because it based on fitting a
 Gaussian-uniform mixture model to the dataset.
 This can be thought of as a simpler version of the spatial plus
 proper motions space modelization found in previous works, for
 example, ~\citet{Jones_1988}.

 A $D$-dimensional Gaussian distribution can be written as

 \begin{equation}
 \mathcal{N}(\mathbf{x} | \mu, \Sigma)=
 \frac{1}{(2 \pi)^{D / 2}|\Sigma|^{1 / 2}}
 \exp \left(-\frac{1}{2}(\mathbf{x}-\mu)^{T} \Sigma^{-1}(\mathbf{x}-\mu)\right),
 \end{equation}

 \noindent where $\mathbf{x}$ is the $D$-dimensional data vector, and $
 (\mu,\Sigma)$ are the mean and covariance matrix.
 A GMM with $K$ components (i.e., Gaussians) is defined as

 \begin{equation}
 \rho_{GMM}=\sum_{i=1}^{K} \pi_{i}
 \mathcal{N}\left(\mathbf{x} | \mu_{i}, \Sigma_{i}\right),
 \end{equation}

 \noindent where $\pi_i$ are the weights (or mixing coefficients)
 associated with each of the $K$ components. Similar to the GMM, we define the
 GUMM as a two-dimensional mixture model composed of a Gaussian, representing
 the stellar cluster, and a uniform distribution, representing the noise due
 to contaminating field stars. The full model is then written as

 \begin{equation}
 \rho_{GUMM}=
 \pi_{0} \mathcal{N}\left(\mathbf{x} | \mu, \Sigma\right)+ \pi_{1} U[0,1],
 \end{equation}

 \noindent where $U[0,1]$ is the uniform distribution in the range $[0,1]$,
 and $\pi_{x} (x=0,1)$ are the unknown weights for each model.
 No restrictions are imposed on the position, shape, or
 extension of the 2D Gaussian representing the stellar cluster.
 Following the recipe employed by the classic GMM, we use the
 iterative expectation-maximization algorithm~\citep[EM,][]{dempster_1977} to
 estimate these  weights as well as the mean and covariance of the 2D
 Gaussian. After the EM algorithm converges to a solution, each star is
 assigned a probability of belonging to the 2D Gaussian 
 (i.e., to the putative cluster). We then need to decide which stars to reject
 as field stars based on these probability values. To do this the percentile
 distribution of the probabilities is generated and the value
 at which the curve begins a sharp climb toward large probabilities is
 automatically identified as the probability cut. The value
 corresponding to the climb in the percentile curve is estimated with the
 method developed in the \texttt{kneebow} package\footnote{
 \url{https://github.com/georg-un/kneebow}}. The user can input a
 manual value for this probability cut (or even skip the GUMM altogether), but
 after extensive testing this method has proven to give very good results and
 it is thus the recommended default.
 Stars below this value are rejected as contaminating field stars and the
 surviving stars are kept as cluster members.

 \begin{figure}
 \includegraphics[width=\hsize]{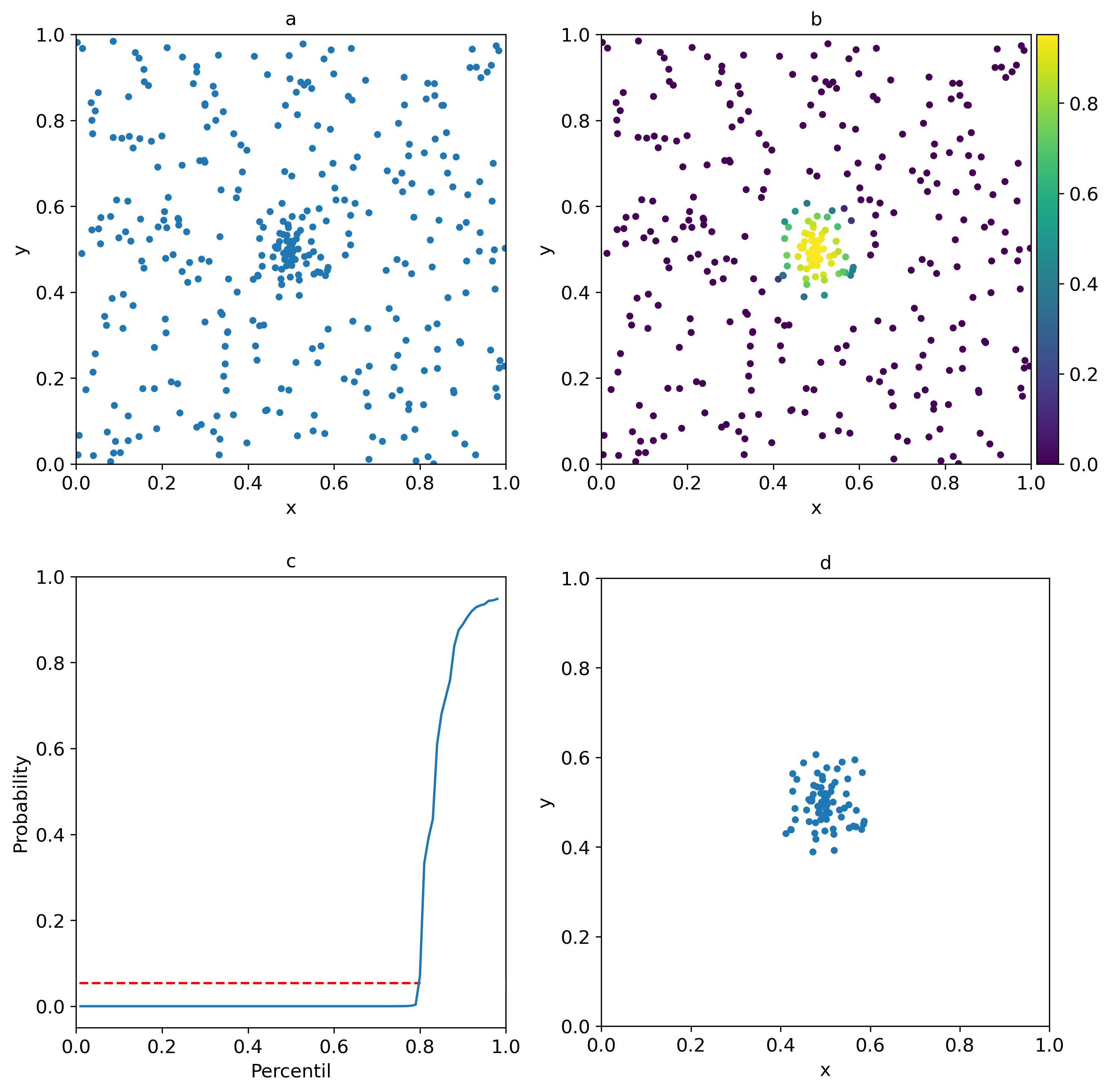}
 \caption{GUMM process in four steps. Panel a: The set of stars that survived
 the RFR block. Panel b: Probabilities assigned by the GUMM to all the stars in the
 frame. Panel c: The method for selecting the probability cut value using a
 percentile plot. Panel d: The final set cleaned from most of the contaminating
 field stars.}
 \label{fig:gumm}
 \end{figure}

 The results of processing a group of stars from a synthetic cluster with the
 GUMM can be seen in Fig.~\ref{fig:gumm}.
 The plot in panel a shows the 2D coordinates space after the RFR
 block rejects those clusters; this is consistent with a random uniform distribution.
 It can be seen that, even after clusters mainly composed of field stars are
 rejected, the central overdensity is still visibly contaminated by the
 surrounding field stars. The plot in panel b shows the probabilities assigned to each
 star of belonging to the 2D Gaussian via the GUMM process. In the plot in panel c, we show
 the percentile diagram for the probabilities, where the red line shows the
 value at which the cut is imposed. Finally, the plot in panel d shows the region after those
 stars with probabilities below the aforementioned cut are rejected.\\

 This process, although almost trivial at first glance, greatly improves the
 purity of the final sample of estimated true members at very little cost
 regarding completeness. The hypothesis at work is of course that the putative
 stellar cluster is more concentrated in the coordinates space than regular
 field stars, as previously stated.

\subsubsection{Kernel density estimator probabilities}
 \label{sssec:kde-probs}

 Once a run of the inner loop is finished, each star in the observed field is
 classified to be either a cluster member or a field star. 
 Although continuous (spatial) probabilities are assigned in the GUMM
 step, these are used to apply a coarse classification between members and
 nonmembers. The information that moves on to the next segment is the hard
 binary classification. This means that only probability values of 0 and 1 are
 assigned up to this stage.
 The KDE block takes these binary probabilities and turns them into
 continuous probabilities in the range [0, 1]. This improves the final results
 in general by assigning more realistic probability values. Furthermore, this
 block is essential for single-run clustering methods (defined in
 Sect~\ref{sssec:clustering}). Clustering methods such as KMS or GMM
 require multiple outer loop runs. The final probabilities are then estimated
 by averaging all the binary probability values, which breaks the binarity.
 Single-run methods work, as the name indicates, on a single run of the outer
 loop. This means that without this block, single-run methods would assign
 probabilities of 0 and 1 exclusively.\\

 The KDE probabilities are assigned after a full run of the inner loop, with
 all stars classified as either members or nonmembers. The process is as
 follows:

 \begin{enumerate}
  \item Separates each of those two classes into different sets.
  \item Estimate the KDE for each class, using all the available data, that is,
  coordinates plus the data dimensions used for clustering (photometry, proper
  motions, etc.).
  \item Evaluate all the data in the frame in the KDE obtained for each class.
  \item Use the above evaluations as likelihood estimates in the Bayesian
  probability for two exclusive and exhaustive hypotheses (i.e., a star
  belongs to either the members distribution or the field stars distribution).
 \end{enumerate}

\begin{figure}
  \includegraphics[width=\hsize]{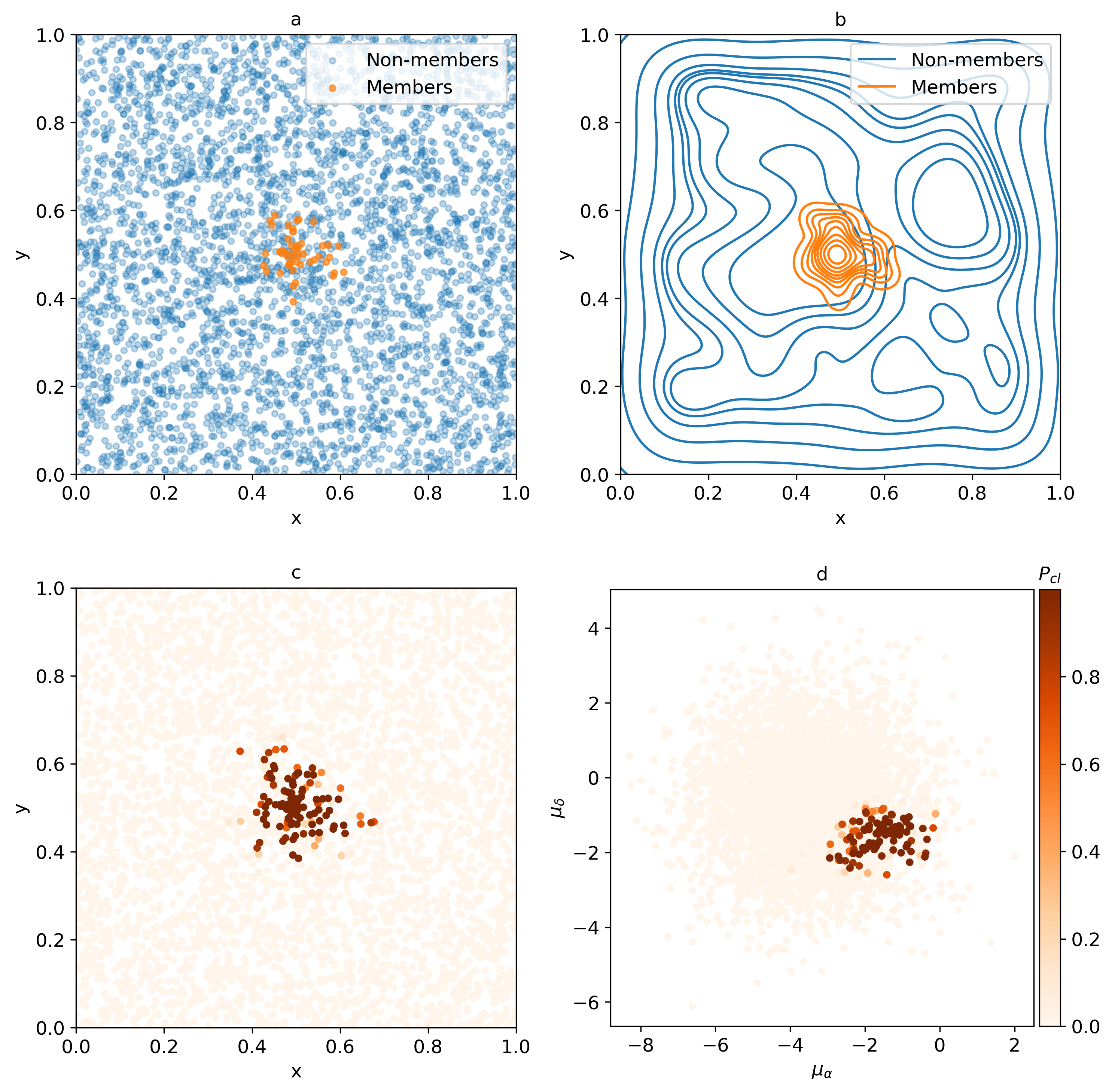}
  \caption{KDE probabilities method shown in the coordinates space. Panel a:
  Members and nonmembers, as estimated by the inner loop process. Panel b:
  KDEs for both classes. Panel c: Final $P_{cl} $ probabilities assigned in the
  coordinates space. Panel d: Same as panel c, but for proper motions.}
  \label{fig:KDE}
  \end{figure}

 The final cluster membership probability (using uniform equal priors) is
 written as

 \begin{equation}
 P_{cl} = KDE_{m} / (KDE_{m} + KDE_{nm}),
 \end{equation}

 \noindent where $KDE_{m}$ and $KDE_{nm}$ are the KDE likelihoods for the
 members and nonmembers (field), respectively. The process can be seen in
 Fig.~\ref{fig:KDE} for the coordinates dimensions (even though it is applied
 on all the data dimensions, described in Sect.~\ref{ssec:synthetic}).
 The plot in panel a shows the two classes, members and
 nonmembers, generated after the inner loop is finished. In the plot in panel b,
 we show the two-dimensional coordinates KDEs for both classes, noting again that this
 is applied on all the data dimensions. The plot in panel c shows the
 nonbinary $P_{cl} $ probabilities assigned by the method in the coordinates
 space. Finally, the plot in panel d is equivalent to the plot in panel c, but for the proper
 motions space.
 
\section{Validation of the method}
 \label{sec:validation}

 In order to perform a thorough comparison of the performance of pyUPMASK
 with that of UPMASK, we applied both methods to a large number of synthetic
 clusters and quantified the results using numerous statistical metrics.
 In this section, we describe the set of synthetic clusters, the selected metrics, and the
 reasoning behind the choice of input parameters.

\subsection{Synthetic datasets}
 \label{ssec:synthetic}

 We employed a total of 600 synthetic clusters to analyze the performance of
 UPMASK and pyUPMASK, the latter in the six configurations mentioned in
 Sect~\ref{sssec:clustering}. This set is divided into a subset of 320
 clusters, and another of 280 clusters. The first subset is equivalent to
 that used in the original UPMASK article (KMM14) in the sense that
 it is composed of clusters with synthetic photometry generated with the same
 process as that used in KMM14. We refer to this subset as PHOT
 hereinafter.
 The second subset contains 280 clusters generated by adding synthetic proper
 motions to all the stars in the frame; we refer to this subset
 as PM hereinafter.
 The idea is to see how the two algorithms handle the case in which only
 photometry is available (i.e., the PHOT dataset), and the
 increasingly common case (thanks to the Gaia mission) in which proper motions
 with very reasonable quality are available (i.e., the PM dataset).
 The performance of UPMASK and pyUPMASK is tested using the 600
 synthetic clusters obtained by combining the PHOT and PM datasets.
 Clusters were generated with a wide range of field star contamination.
 The level of contamination is measured by the contamination index (CI),
  which is defined as the number of field stars to cluster members in the frame 
 to match the ``contamination rate'' used in KMM14. The
 maximum CI in our set of synthetic clusters is 200.

\begin{figure}
 \includegraphics[width=\hsize]{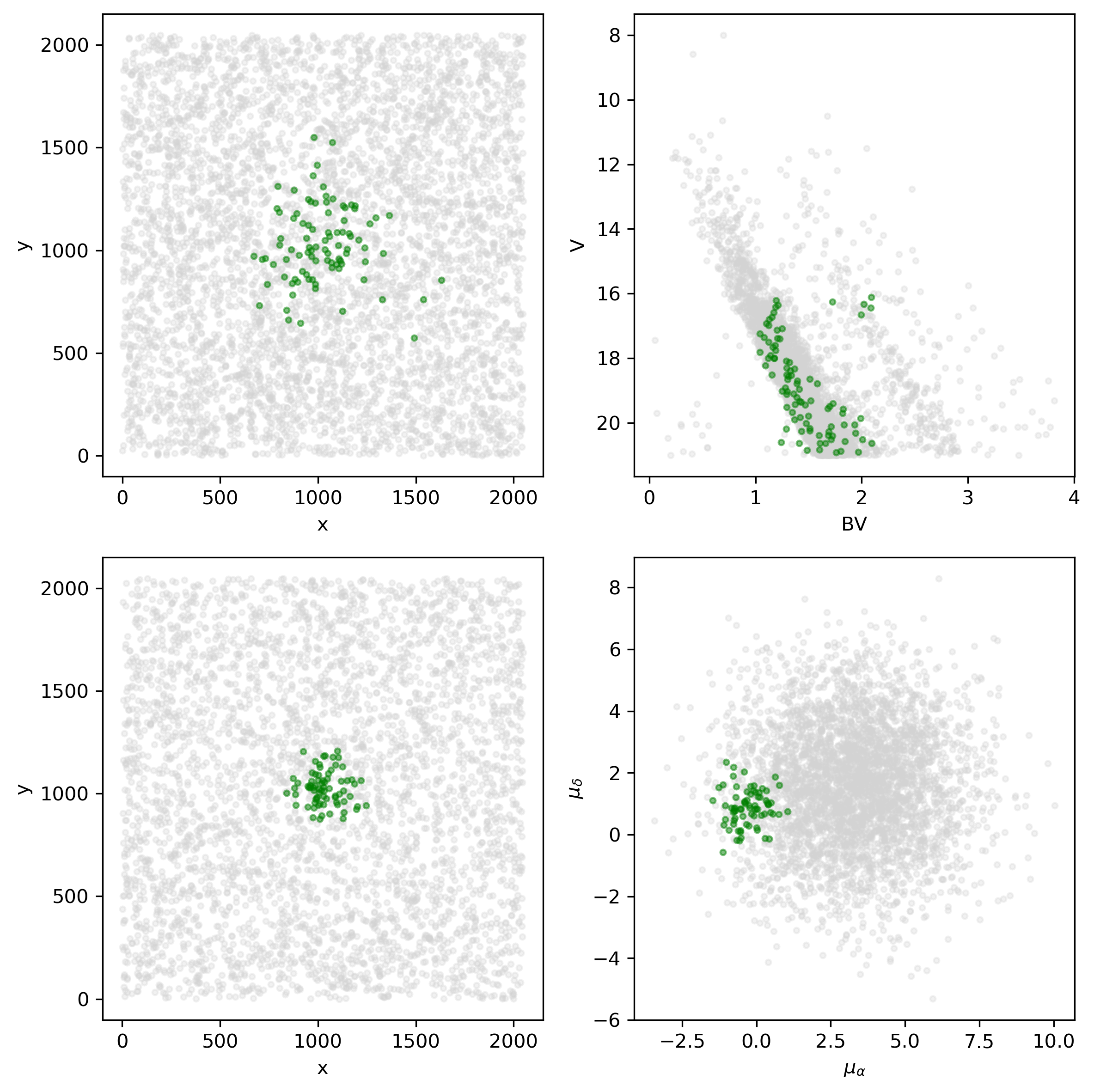}
 \caption{Top row: Coordinates and CMD for a PHOT synthetic
 cluster with moderate CI.
 Bottom row: Coordinates and vector-point diagram for a PM synthetic
 cluster with moderate CI.}
 \label{fig:synth_clust}
 \end{figure}

 In Fig.~\ref{fig:synth_clust} we show examples of a PHOT (top) and PM
 (bottom) synthetic clusters, which are generated with moderate contamination 
 (CI$\approx$50).

\subsection{Performance metrics}
 \label{ssec:performance}



 %
 %
 A proper choice for evaluating the classification performance of a
 probabilistic model (such as UPMASK or pyUPMASK) is a debate that carries on
 even today~\citep{Hand_2009,hernandez_2012}. Different metrics or scoring
 rules yield different results regarding the performance of the model~\citep{Merkle_2013}, which means that relying on a single metric
 is not recommended. This is particularly true when dealing with datasets that
 can be highly imbalanced, as is our case. We thus chose to employ multiple
 metrics. By combining all of these, we expect to obtain a non-biased
 assessment of the overall performance of pyUPMASK versus UPMASK.

 We selected six metrics that can be divided into two groups of three each.
 The first group consists of strictly proper scoring rules, which
 guarantee that they are only optimized when the true classification is
 obtained. This group is composed of the following metrics:\\

 \noindent  Logarithmic scoring rule:
 \begin{equation}
 LSR = 1+\frac{1}{N} \sum_{i=1}^N y_{true} \log(p) + (1-y_{true}) \log(1-p),
 \end{equation}
 \noindent where $N$ is the number of elements, $y_{true}\in\{0,1\}$ is the
 true label, and $p{=}\operatorname{Pr}(y{=}1)$ is the probability that
 $y{=}1$, that is, the probability that the element belongs to the class
 identified with a 1~\citep{Good_1952}. The LSR (also called log-loss or
 cross-entropy) heavily penalizes large differences between $y_{true}$ and
 $p$.\\

 \noindent Brier score loss:
 \begin{equation}
 BSL = 1-\frac{1}{N} \sum_{i=1}^N (p-y_{true})^2,
 \end{equation}
 \noindent which is equivalent to the mean squared error for binary classification;
 it was originally introduced in~\cite{Brier_1950}.\\

 \noindent H measure:
 \begin{equation}
 HMS = 1 - \frac{L}{L_{max}},
 \end{equation}
 \noindent where $L$ is the loss function, and $L_{max}$ is the maximum
 loss; the expression for the loss function is much too mathematically
 involved to be presented here, it can be seen in full in~\cite{Hand_2009}.
 This is a relatively new metric. It was developed as a replacement of the
 popular AUC (area under the receiver operating characteristic curve) score;
 now known to be an incoherent performance measure and thus not
 recommended~\citep{Lobo_2008,Parker_2011,Hand_2014}. The HMS automatically
 handles unbalanced classes by treating the misclassification of the smaller
 class (in our case almost always true members, except for extremely low CI
 values) as more serious than those of the larger class.\\

 \noindent It is worth noting that the definitions of LSR and BSL were altered
 from their original forms by multiplying by -1 and adding plus 1. This way all the
 metrics defined assign 1 to a perfect score.\\

 The three metrics in the first group can be used directly on the membership
 probabilities in the [0, 1] range, resulting from UPMASK or pyUPMASK.
 The second group defined below consists of scoring rules that are applied
 to binary classifiers. These are the types of metrics used in the original
 KMM14 article and we employ them in this work for consistency\footnote{We note that in
 KMM14 the statistical measures TPR and MMR are incorrectly defined. What the
 authors call ``TPR'' is the PPV, and what they call ``MMR''
 is the properly defined TPR.}.
 In the definitions that follow TP is a true positive (a member star correctly
 classified as such), TN is a true negative (a field star correctly classified
 as such), FN is a false negative (a member star incorrectly
 classified as field), and FP is a false positive (a field star incorrectly
 classified as member):\\

 \noindent True positive rate:
 \begin{equation}
 TPR = \frac{TP}{TP+FN},
 \end{equation}
 \noindent which is also called sensitivity or recall; it measures the proportion of
 true members that are correctly identified.\\

 \noindent Positive predictive value:
 \begin{equation}
 PPV = \frac{TP}{TP+FP},
 \end{equation}
 \noindent which is also called precision; it measures how many stars classified as
 members are true members.\\

 \noindent Matthews correlation coefficient:
 \begin{equation}
 MCC = \frac{TP \times TN - FP \times FN}{\sqrt{(TP + FP)(TP + FN)(TN + FP)(TN + FN)}},
 \end{equation}
 \noindent which was introduced in \cite{Matthews_1975}; it can be thought of as an
 equivalent to Pearson's correlation coefficient for binary classifiers. Unlike
 the TPR and PPV, the MCC also takes the TNs into account. It is
 recommended when dealing with imbalanced classes, as is our case.\\

 \noindent To turn the problem into one of binary classification and to be able
 to use the three metrics defined in the second group, we must first select a
 probability threshold that separates the stars into the members and
 nonmembers classes. In KMM14 a single threshold of 90\% was used. Since the choice
 of a threshold can affect the results from these three metrics, we decided
 to use the following two different thresholds: 50\% and 90\%. This way we end up with  the following nine
 metrics to test the performance of UPMASK and pyUPMASK: LSR, BSL, HMS, TPR$_5$,
 PPV$_5$, MCC$_5$, TPR$_9$, PPV$_9$, and MCC$_9$; where the subindex 5 and 9
 indicate the 50\% and 90\% thresholds, respectively.

\subsection{Input parameters selection}
 \label{ssec:input_pars}

 \begin{figure}
 \includegraphics[width=\hsize]{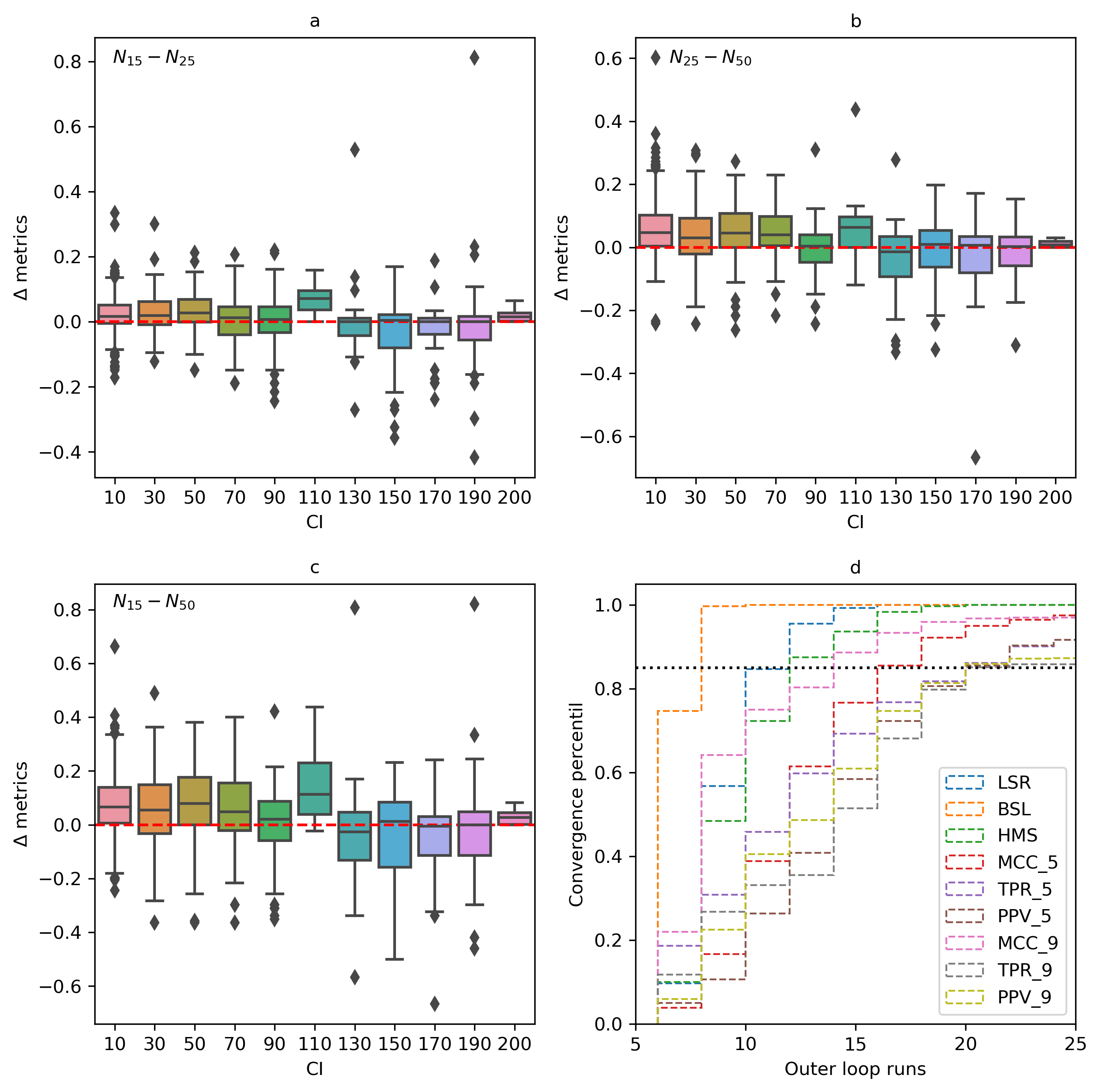}
 \caption{a, b, c: Boxplot of the combined metrics difference vs. CI for the
 100 synthetic clusters used in the test. Combinations for the
 $N_{15},\,N_{25},\,N_{50}$ values are shown.
 Panel d: Outer loop convergence analysis. The convergence percentile
 of the nine metrics vs. the number of outer loop run is shown. The black
 dashed line indicates the 90\% convergence point.}
 \label{fig:UPMASKruns}
 \end{figure}

 There are two main parameters in UPMASK and pyUPMASK that affect the outcome
 of the methods: the number of stars per cluster and the number of runs of the
 outer loop.
 The former, which we refer to as $N_{clust}$, was investigated in KMM14, 
 in which the authors concluded that a value between 10 to 25 is appropriate. In
 the latest version (v1.2) of the UPMASK code, depending on how it is run, the
 default value for $N_{clust}$ is either 25 or 50\footnote{It is 50 if we run
 the code using the \texttt{UPMASKfile} function, and 25 if we use the
 \texttt{UPMASKdata} function.}.
 We performed our own tests using 100 synthetic clusters (50 PHOT and 50 PM)
 covering the full CI range, selected at random from the full list of 600
 mentioned in Sect~\ref{ssec:synthetic}. This set was analyzed with the nine
 performance metrics described in Sect~\ref{ssec:performance}.
 In Fig.~\ref{fig:UPMASKruns} we show the results obtained for three
 $N_{clust}$ values 15, 25, and 50. We combined all the metrics for the 100
 synthetic clusters into one set and subtracted these (900) values for a
 given $N_{clust}$ value from another. The results are plotted versus the CI of
 the synthetic clusters. From panels a to c the combinations
 $N_{15}-N_{25}$, $N_{15}-N_{50}$, and $N_{25}-N_{50}$ are shown, where a
 positive value means that the $N_{clust}$ value on the left performed better
 than the value on the right, and vice versa for negative values. As can be seen,
 the differences are rather small and do not tend to change for different CI
 values.
 We thus decided to use the middle value $N_{clust}=25$ for all the UPMASK and
 pyUPMASK runs, as a reasonable number of default stars per cluster for all the
 CI range.\\

 Deciding how many times the outer loop should run is the other important
 parameter: a low number terminates the code before it is able to present
 fully converged probability values and a large number wastes processing
 time. We processed the same set of 100 synthetic clusters with $N_{clust}=25$
 and analyzed when each of the nine metrics converged to a stable value. The
 stabilization point is defined as the outer loop run where the metric changes
 inside the $\pm0.025$ range for five consecutive runs. The results are shown in
 Fig.~\ref{fig:UPMASKruns} d. plot as a the convergence percentile 
 (i.e., the percentage of clusters that have converged) for each metric versus
 the outer loop run. Almost all the metrics reach a
 convergence above 90\%  before the 25th outer
 loop run. The two exceptions are TPR$_9$ and PPV$_9$, which still show a
 convergence above 85\% before the 25th run.
 Given these results we use 25 runs in the outer loop for all the UPMASK
 and pyUPMASK analyses with the obvious exceptions of the single-run methods
 described in Sect~\ref{sssec:clustering}.


 The PHOT set was processed using all the available photometry as input ($V,
 B-V, U-B, V-I, J-H, H-K$) but selecting only the four principal dimensions
 after the principal component analysis dimensionality reduction.
 For the PM set we used the proper motions ($\mu_{\alpha}, \mu{\delta}$),
 with no dimensionality reduction. Proper motions are generally
 regarded as better cluster members discriminators than photometry 
 owing to the rounded shape of its distribution in contrast with the
 irregular shape of the sequence of a cluster in the photometric space.

\section{Results}
 \label{sec:results}

 To ensure that the results are comparable between the pyUPMASK and UPMASK runs,
 all the analyses were performed on the same computer cluster. In what follows, the
 results are classified according to whether pyUPMASK or UPMASK performed
 better for a given metric and synthetic cluster.
 We allow for a small range of $\pm0.005$ to act as a ``tie zone''
 in which the two methods can be thought of as performing equally well.
 In Appendix~\ref{apx:results} we show the results of comparing
 pyUPMASK with the Bayesian method included in \texttt{AsteCA}. These are not
 included here because the methods are not directly comparable, as explained in
 the Appendix.\\


 \begin{figure*}
 \resizebox{\hsize}{!}{\includegraphics{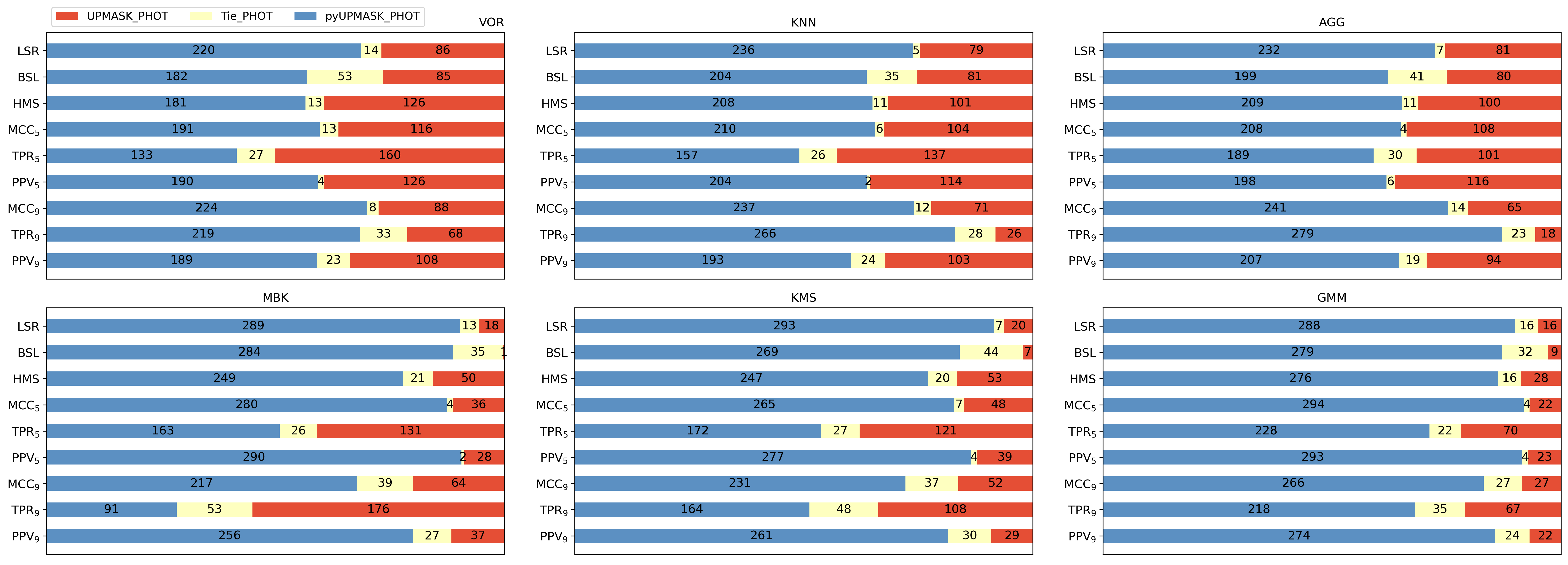}}
 \caption{Results for the 320 synthetic clusters in the PHOT dataset
 processed with the six clustering methods used in pyUPMASK vs. UPMASK. For
 each metric, the blue and red bars represent the cases where pyUPMASK and UPMASK
 performed better, respectively. The yellow bars represent cases in which both
 methods performed equally well.}
 \label{fig:allmethods_PHOT}
 \end{figure*}

 \begin{figure*}
 \resizebox{\hsize}{!}{\includegraphics{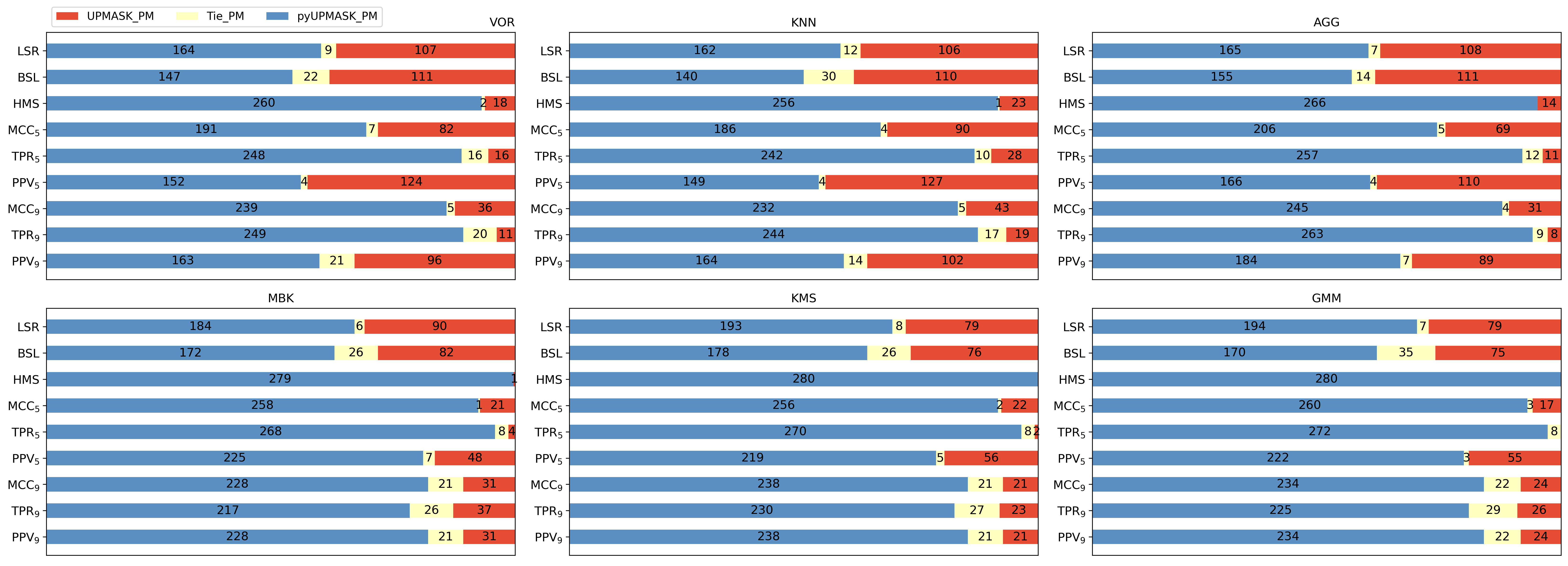}}
 \caption{Same as Fig.~\ref{fig:allmethods_PHOT} but for the 280 synthetic
 clusters in the PM dataset, exclusively.}
 \label{fig:allmethods_PM}
 \end{figure*}

 In Figs.~\ref{fig:allmethods_PHOT} and~\ref{fig:allmethods_PM} we show
 the metrics for the 320 and 280 synthetic clusters in the PHOT and PM
 datasets, respectively, for each of the six clustering methods used in
 pyUPMASK. The blue, yellow, and red bars depict the proportion of cases for which
 pyUPMASK performed better, equally well, and for which UPMASK performed
 better, respectively. It is easy to see that, although with some variation
 across clustering methods, pyUPMASK has a better performance than UPMASK for
 all the methods and all the metrics, particularly for the PM
 dataset. This is an outstanding result that unmistakably shows the large
 improvement  brought by pyUPMASK.
 The three methods that apply multiple outer loop runs (MBK, KMS, GMM) show a
 clear advantage over the remaining single-run methods, regarding the
 proportion of cases for which pyUPMASK resulted in larger metric values.\\

 In~\cite{Cantat_2018} the authors used a modified version of UPMASK to
 estimate membership probabilities for more than 1200 cataloged clusters. The
 modification was motivated by the need to increase the speed for processing
 large numbers of clusters. This modification mainly consists in replacing the default KDE based method in the RFR block in UPMASK for a minimum spanning tree method 
 (see article for more details).
 We tested this modified version\footnote{Thanks to Dr Cantat-Gaudin
 who shared the code with us.}, which we refer to as MST, using the same
 set of synthetic clusters and metrics employed so far. The code was executed
 with 25 runs of the outer loop and 15 stars per cluster; internal tests
 showed that this gave more adequate results than using 25.\\

\begin{figure}
 \includegraphics[width=\hsize]{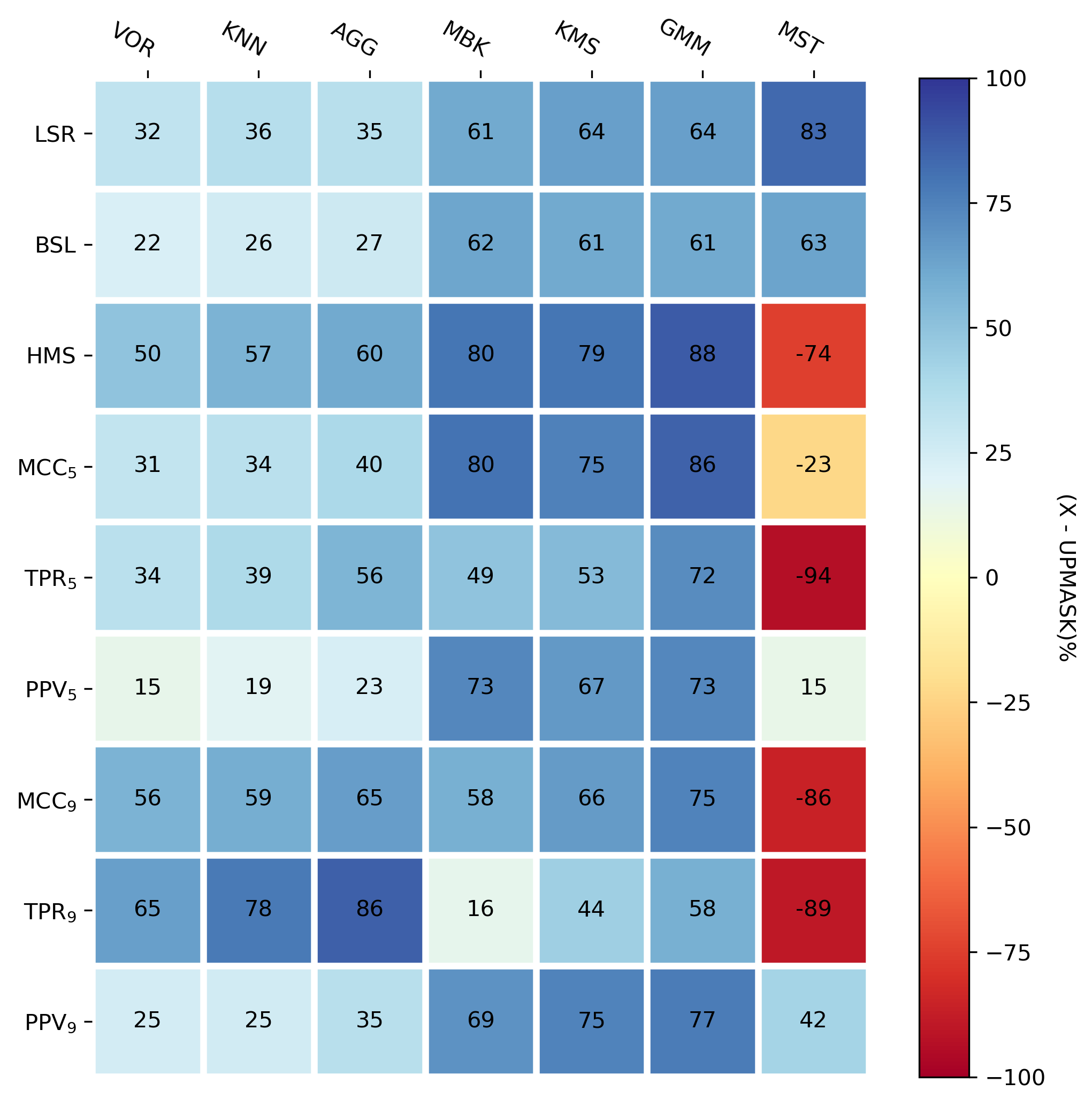}
 \caption{Matrix plot of the six pyUPMASK clustering methods plus MST, vs.
 the nine defined metrics for the 600 synthetic clusters. Each square shows
 the percentage difference of the number of cases for which pyUPMASK/MST performed
 better, minus the number for which UPMASK performed better.}
 \label{fig:matrix}
 \end{figure}

 The results of our six clustering methods, plus the MST method, versus UPMASK
 can be compressed into a single matrix plot as shown in
 Fig.~\ref{fig:matrix}. We show the X minus UPMASK percentage
 metric difference, where X represents each of the pyUPMASK clustering methods
 plus MST. This value is obtained subtracting the number of synthetic clusters,
 where pyUPMASK/MST performed better than UPMASK, from the number of clusters
 where UPMASK showed a better performance, and taking the percentage.
 This difference ranges from -100, which would indicate that UPMASK performed
 better on all 600 synthetic clusters, to 100, indicating that pyUPMASK (or
 MST) was the better performer for the 600 clusters. A value of 0 indicates
 that both methods performed better on an equal number of cases.
 As can be seen, for the pyUPMASK methods all the squares in the matrix are
 positive (the smallest being the PPV$_5$ metric for the VOR method), which
 again shows that pyUPMASK performed significantly better than UPMASK, measured
 by any of the employed metrics.
 The advantage of the MBK, KMS and GMM methods over the single-run methods is
 easier to see here compared to Figs.~\ref{fig:allmethods_PHOT}
 and~\ref{fig:allmethods_PM}. The only exception
 is the TPR$_9$ metric for which the VOR, KNN, and AGG methods show a larger
 differential than the remaining multiple-runs methods; that is, more true members
 are classified as such. This comes at the expense of the PPV$_9$ metric, for which
 the MBK, KMS and GMM methods show much larger values; that is, fewer field stars are
 incorrectly classified as members. Other than this, there is no visible
 relation between any clustering method and a given performance metric.

 The MST method shows a somewhat erratic behavior across the metrics. It
 performs worse than UPMASK for almost all of the clusters for several metrics
 (i.e., HMS, TPR$_5$, MCC$_9$ and TPR$_9$) and better for a few others (e.g. LSR and BSL).
 Overall, the statistical performance of the MST method is worse than
 UPMASK and pyUPMASK with any of the tested clustering methods.
 Notwithstanding, MST is faster than UPMASK (as we show below)
 and outperforms all other methods in the LSR and BSL metrics.\\

\begin{figure*}
 \resizebox{\hsize}{!}{\includegraphics{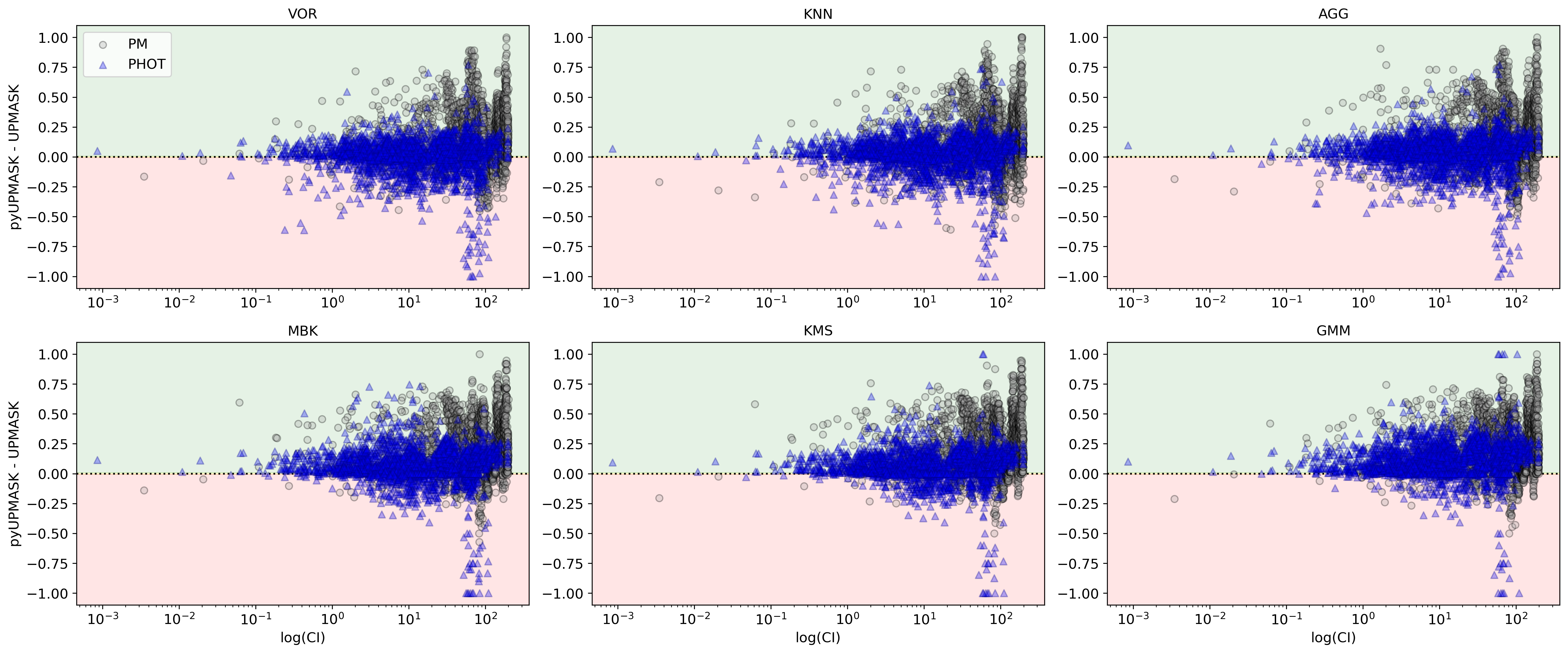}}
 \caption{Differences between pyUPMASK vs. UPMASK results for all the
 metrics combined, vs. the CI (shown as a logarithm). Each clustering method
 is shown separately, as are the PHOT and PM sets using blue triangles and
 black circles, respectively. The red and green regions correspond to the
 regions for which pyUPMASK gives worse and better results than UPMASK,
 respectively.}
 \label{fig:CIdelta}
 \end{figure*}

 In Fig.~\ref{fig:CIdelta} we show the dependence with CI for the pyUPMASK
 minus UPMASK difference for all the metrics, for each clustering method. A
 positive value (green region) means that pyUPMASK performed better, while a
 negative value (red region) means that it performed worse than UPMASK. The
 PHOT and PM sets are shown with triangles and circles, respectively.
 There is no apparent trend with CI for the results of any clustering method.
 What is clear is that pyUPMASK performs even better for the PM set as
 evidenced by the overall larger (more positive) differences, particularly for
 clusters with large CI values. This is a very desirable result taking into
 account that high quality proper motions are becoming more accessible very
 fast.\\

 \begin{table}
 \caption{Aggregated results for all the metrics and all the synthetic
 clusters, for the six pyUPMASK clustering methods used, as percentage of
 results where pyUPMASK outperformed UPMASK, and vice versa, respectively. The
 missing percentage to add up to 100 corresponds to ties.
 The second rows for each method show the minimum and maximum percentage values
 obtained for any single metric (shown in parenthesis), for that method.}
 \label{tab:results}
 \centering
 \begin{tabular}{l c c l}
 \hline\hline
 Method & pyUPMASK & UPMASK\\
  & min | max & min | max \\
 \hline
   VOR & 66 & 29 \\
   & 55 (BSL) | 78 (TPR$_9$) & 13 (TPR$_9$) | 42 (PPV$_5$)\\
   KNN & 68 & 27\\
   & 57 (BSL) | 85 (TPR$_9$) & 8 (TPR$_9$) | 40 (PPV$_5$)\\
   AGG & 72 & 24 \\
   & 59 (BSL) | 90 (TPR$_9$) & 4 (TPR$_9$) | 38 (PPV$_5$)\\
   MBK & 77 & 16\\
   & 51 (TPR$_9$) | 90 (MCC$_5$) & 8 (HMS) | 36 (TPR$_9$)\\
   KMS & 79 & 15\\
   & 66 (TPR$_9$) | 88 (HMS) & 8 (PPV$_9$) | 22 (TPR$_9$)\\
   GMM & 83 & 11\\
   & 74 (TPR$_9$) | 93 (HMS) & 5 (HMS) | 16 (LSR)\\
 \hline
 \end{tabular}
 \end{table}

 We can further compress the results by combining each metric into a single
 value, for each of the clustering methods tested in pyUPMASK. That is, we take
 the 5400 results for each clustering method (600 synthetic clusters times nine
 metrics) and calculate the percentage at which pyUPMASK outperformed UPMASK.
 The same process can be applied to the synthetic clusters for which
 UPMASK outperformed pyUPMASK to obtain a similar, inverted, percentage.
 The results are shown in Table~\ref{tab:results}.
 This table shows that even the worst pyUPMASK performer (VOR) gives
 better metrics than UPMASK 66\% of the times. The method  with the highest
 pyUPMASK percentage (GMM) outperforms UPMASK 83\% of the times, which is a massive
 advantage. The worst individual metric result is obtained for TPR$_9$ in the
 MBK method. Still the value is larger than 50\%, which means that the majority
 of the cases were better handled by pyUPMASK. On the other end of the
 analysis the best metric result is found for HMS in the GMM method, for which
 pyUPMASK manages to outperform UPMASK for virtually all of the cases.\\

\begin{figure}
  \centering
  \includegraphics[width=\hsize]{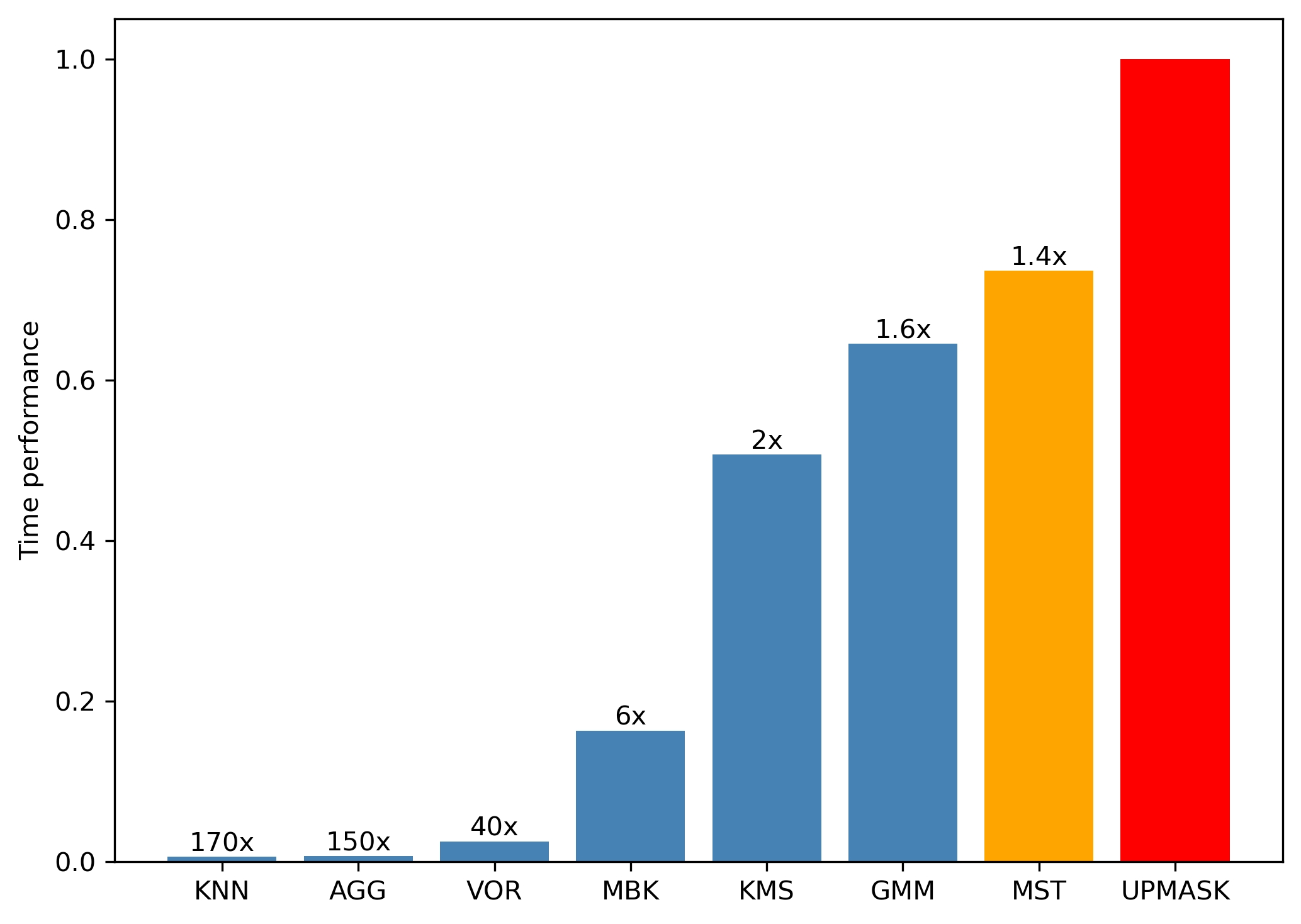}
  \caption{Amount of time employed in processing the 600 synthetic
  clusters by each pyUPMASK method (blue bars), the MST method (orange bar),
  and UPMASK (red bar). The bars are normalized so that UPMASK corresponds to a
  total value of 1.}
  \label{fig:timeanalys}
 \end{figure}

 Another important aspect along with the performance measured by the
 statistical metrics is the performance measured in computing time. This is
 shown in Fig.~\ref{fig:timeanalys} as a bar plot normalized to the total time
 used by UPMASK to process the 600 synthetic clusters. The numbers on top of
 the bars displays how many times faster each clustering method in pyUPMASK
 is compared to UPMASK. We also show the time
 performance of the MST modification mentioned previously.
 The fastest method is expectedly a single-run method, KNN,
 which performs 170 times faster than UPMASK. This is an enormous margin of
 difference. Even the slowest method, GMM, is faster than UPMASK: this method manages to
 process the set of synthetic cluster employing almost 38\% less time than
 UPMASK or 1.6 times faster. On average we can say that pyUPMASK using a
 single-run method is over 100 times faster than UPMASK and is more than 3 times
 faster for the multiple run methods.\\

 The choice between which clustering method to employ in pyUPMASK then
 depends on the specific requirements of the analysis. If the absolute best
 performance measured by a classification metric is sought after, then clearly
 GMM is the method to chose (with the advantage of being faster than
 UPMASK). If we can trade some performance for a faster
 process, then KMS or MBK can be used. And if we are willing to trade even more
 classification performance, while still performing much better than UPMASK,
 then VOR, KNN, or AGG are by far the fastest approaches.\\


 Finally, we consider the issue of computational resources requirements.
 We found that for very large input data files memory and
 processing power requirements can be too much for most methods to handle.
 Although the VOR clustering method is the worst performer out of the six
 tested methods (measured by classification metrics), it has an advantage
 compared to all the others, including UPMASK, when it comes to analyzing large
 files. 
 To obtain the Voronoi diagram of an N-dimensional set of points, the Python 
 \texttt{scipy} package relies on the Qhull
 library~\citep{Barber_1996}\footnote{\url{http://www.qhull.org/}}.
 This library is written in the C language which makes it extremely efficient,
 thus making the pyUPMASK VOR method very efficient for large datasets.

 To test this we downloaded a large 6$\times$6 deg region around the NGC2516
 cluster from the Gaia second data release~\citep{GaiaDR2_2018}. The resulting
 field contains over 420000 stars up to a maximum magnitude of $G=19$ mag. This
 limit was imposed because beyond this value the photometric errors grow
 exponentially.
 The frame was processed with the six tested pyUPMASK clustering methods
 and UPMASK, using proper motions and parallax as input data. We used 25
 outer loop iterations for all the methods, except of course for
 the single-run methods, and a value of 25 for the parameter that determines
 the number of elements per cluster (i.e., the default values for both
 parameters as explained in Sect.~\ref{ssec:input_pars}).\\

 \begin{figure*}
 \resizebox{\hsize}{!}{\includegraphics{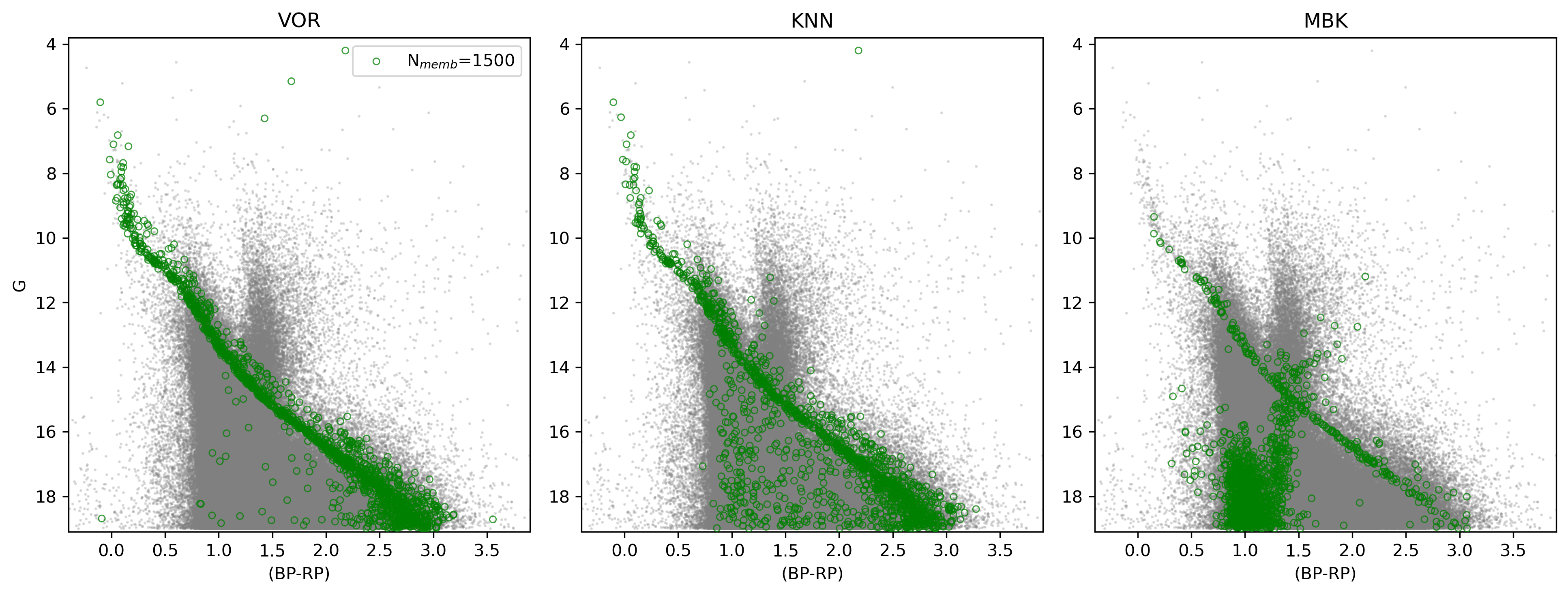}}
 \caption{Results for the NGC2516 cluster by the VOR (left), KNN (center), and
 MBK (right) methods. Estimated members are shown as green circles; the field stars
 are shown as gray dots.}
 \label{fig:NGC2516}
 \end{figure*}

 Only three methods were able to complete the process: VOR, KNN, and MBK. The
 methods AGG, KMS, and GMM failed owing to memory requirements as they
 attempted to allocate arrays of $\sim$640 Gb, $\sim$31 Gb, and
 $\sim$31 Gb on memory, respectively. The UPMASK algorithm was not able to finish even the
 first iteration of the inner loop within the first iteration of the outer loop
 after a full week of running, so it was halted.
 The results of the VOR, KNN, and MBK methods can be seen in a color-magnitude
 diagram (CMD) in Fig.~\ref{fig:NGC2516}. We plotted the 1500 stars with larger
 membership probabilities given by each method, as this is the approximate
 number of cluster members in the frame (given by a simple stellar density
 analysis). It is evident that the VOR method returns the most reasonable and
 less contaminated CMD out of the three. Furthermore, this method managed to
 process the cluster almost 4 and 40 times faster than KNN and MBK,
 respectively. It is worth noting that on a personal computer, which has far
 less resources than a computational cluster, VOR was the only method that
 was able to run.\\

 A smaller field containing this same cluster was analyzed with UPMASK
 in~\cite{Cantat_2018}. The processed area contains only
 $\sim$1100 stars associated with this cluster up to a magnitude of G${=}18$ mag.
 The analysis done in this work resulted in less than 800 stars
 with membership probabilities (MPs) above 0.5 and $\sim$100 stars with
 MPs$>0.9$. In contrast, using the same magnitude cut, we are able to obtain
 with the VOR method on our very large field over 1700 stars with MPs$>0.99$
 tracing a well-defined sequence. The advantage of studying a cluster
 using almost all of its members versus using less than 10\% of the members 
 (comparing the large MPs subsets), is obvious.\\

 The VOR method is thus the only one that was able to produce quality results
 for this very large dataset, and it did so while using the least amount of
 processing time by a wide margin.



\section{Conclusions}
 \label{sec:conclusions}

 Since its development in KMM14 the UPMASK code has been used to analyze
 thousands of clusters. This is because it is a very
 smart, general, and efficient unsupervised method, that requires no prior
 knowledge about the observed field.
 In this work we introduced pyUPMASK, a tool based on the general UPMASK
 algorithm with several added enhancements. The primary aim of pyUPMASK is the
 assignment of membership probabilities to cluster stars observed in a frame
 contaminated by field stars.
 We tested our code extensively using 600 synthetic clusters affected
 by a large range of contamination. Six performance metrics were employed,
 three of which were in two different configurations, to ensure sufficient coverage
 when assessing statistical classification.
 The results from six different clustering methods in pyUPMASK were compared to
 those from UPMASK.
 Under the conditions established for the analysis, the pyUPMASK tool proved to
 clearly outperform UPMASK 
 while still managing to be faster (and, for the single-run methods, extremely
 faster).

 This new tool is thus highly configurable (around a dozen clustering
 algorithms supported), fast, and an excellent performer measured by
 several metrics.
 The pyUPMASK algorithm is fully written in Python and is made available for its use
 under a GPL v3 general public
 license\footnote{\url{https://www.gnu.org/copyleft/gpl.html}}.

\begin{acknowledgements}
The authors would like to thank the anonymous referee, for their
helpful suggestions and corrections to the manuscript.
The authors would like to thank Dr Anagnostopoulos for his help with the
\texttt{hmeasure} (\url{https://github.com/cran/hmeasure}) R package.
The authors would like to thank Dr Cantat-Gaudin for sharing the code used
in~\cite{Cantat_2018}.
M.S.P., G.I.P., and R.A.V. acknowledge the financial support from CONICET 
(PIP317) and the UNLP (PID-G148 project).
AM acknowledges support from the Portuguese Fundação para a Ciência e a
Tecnologia (FCT) through the Strategic Programme UID/FIS/00099/2019 for CENTRA.
This research has made use of NASA's Astrophysics Data System.
This research made use of the Python language~\citep{vanRossum_1995}
and the following packages:
NumPy\footnote{\url{http://www.numpy.org/}}~\citep{vanDerWalt_2011};
SciPy\footnote{\url{http://www.scipy.org/}}~\citep{Jones_2001};
Astropy\footnote{\url{http://www.astropy.org/}}, a community-developed core Python
package for Astronomy \citep{astropy:2013,astropy:2018};
scikit-learn\footnote{\url{http://scikit-learn.org/}}~\citep{pedregosa_2011};
matplotlib\footnote{\url{http://matplotlib.org/}}~\citep{hunter_2007}.
This work has made use of data from the European Space Agency (ESA) mission
{\it Gaia} (\url{https://www.cosmos.esa.int/gaia}), processed by the {\it Gaia}
Data Processing and Analysis Consortium (DPAC,
\url{https://www.cosmos.esa.int/web/gaia/dpac/consortium}). Funding for the DPAC
has been provided by national institutions, in particular the institutions
participating in the {\it Gaia} Multilateral Agreement.
\end{acknowledgements}

\bibliographystyle{aa}
\bibliography{biblio}

\begin{appendix}
\section{pyUPMASK versus \texttt{ASteCA} results}
 \label{apx:results}

We present a comparison between the membership probability estimation
algorithm included in \texttt{ASteCA} and pyUPMASK. It is worth noting that
\texttt{ASteCA} is a complete package for processing stellar clusters that
includes a Bayesian membership estimation method. This method,
which has not changed since the \citet{Perren_2015} article was published, is
based on comparing  the distributions of field stars and stars within the
cluster region in whatever data space the user decides to use (photometric,
proper motions, parallax, or any combination). The cluster region is defined by
the center coordinates and radius values estimated by separate methods in 
\texttt{ASteCA} that were applied previous to the Bayesian membership method.
The pyUPMASK method (similarly UPMASK), on the other hand, is a method for estimating
membership probabilities. That is, it represents just a portion of what the 
\texttt{ASteCA} package comprises.

The reason for not including this comparison in the main article is that the
Bayesian method in \texttt{ASteCA} and pyUPMASK are not directly comparable.
Unlike UPMASK and pyUPMASK, which are unsupervised methods, the membership
method included in \texttt{ASteCA} is supervised because it requires an a
priori separation of classes. That is: the field stars, identified as those
stars located in the field region, and the possible cluster members, identified
as those stars located in the cluster region, must be segregated before the
membership method can be applied. Hence, the membership probabilities obtained
with the Bayesian method in \texttt{ASteCA} are a reflection not only of
the method itself, but also of the separate methods used to estimate the
center and radius values.
\\

\begin{figure*}
 \resizebox{\hsize}{!}{\includegraphics{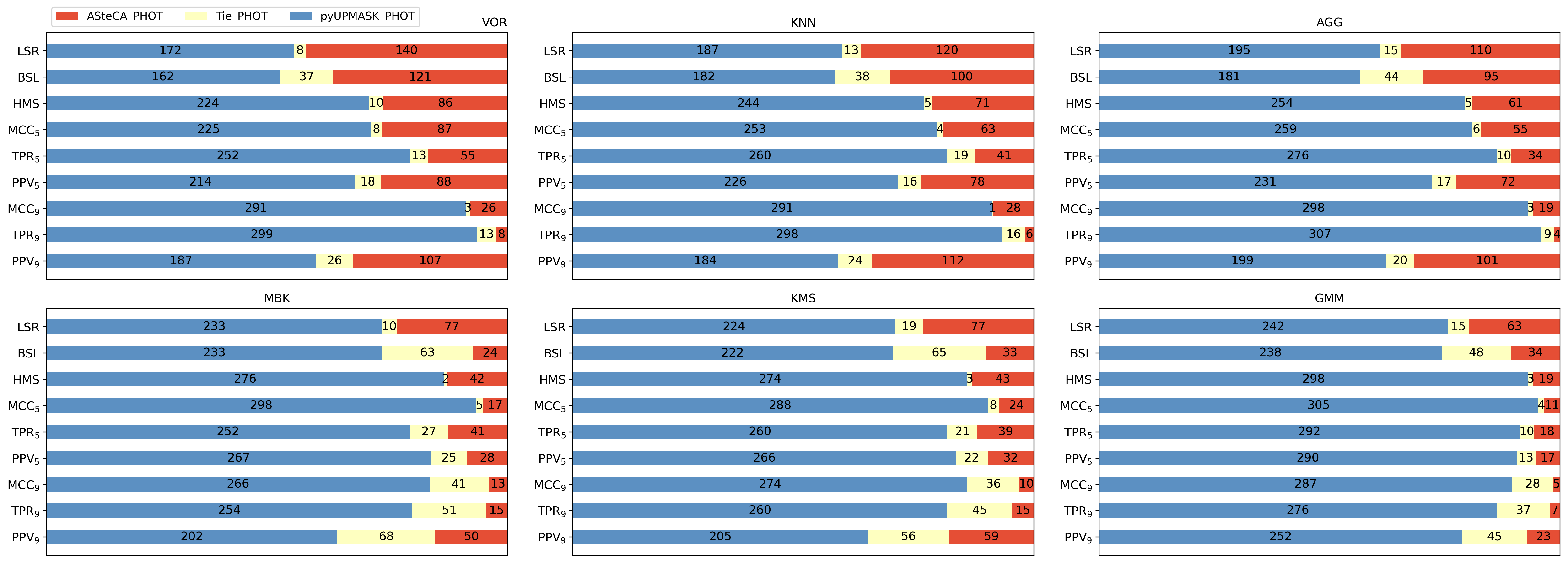}}
 \caption{Same as Fig.~\ref{fig:allmethods_PHOT} but showing pyUPMASK versus 
 \texttt{ASteCA}.}
 \label{fig:allmethods_PHOT_asteca}
 \end{figure*}

\begin{figure*}
 \resizebox{\hsize}{!}{\includegraphics{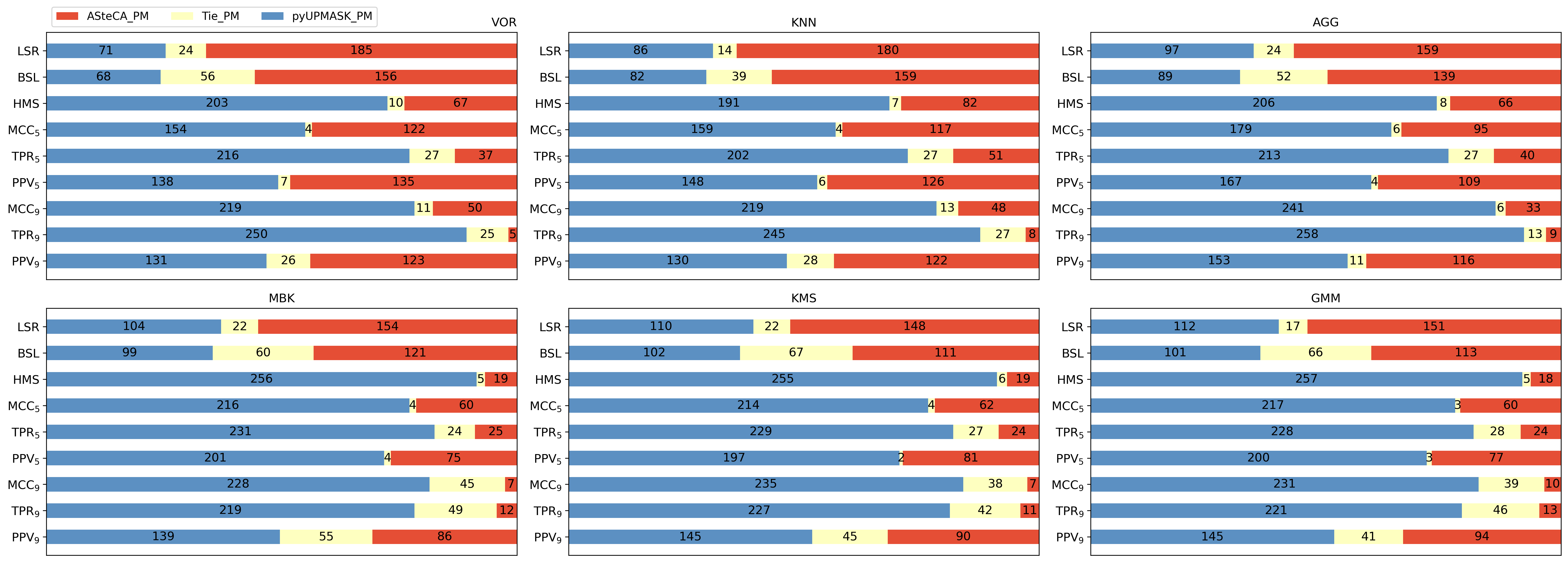}}
 \caption{Same as Fig.~\ref{fig:allmethods_PM} but showing pyUPMASK versus 
 \texttt{ASteCA}.}
 \label{fig:allmethods_PM_asteca}
 \end{figure*}

The \texttt{ASteCA} algorithm was thus applied on both datasets (PHOT and PM), allowing it to
automatically estimate the center coordinates and radius
value of the synthetic cluster. As shown in Figs~\ref{fig:allmethods_PHOT_asteca}
and~\ref{fig:allmethods_PM_asteca}, pyUPMASK performs better than
\texttt{ASteCA} for both datasets, particularly for the PHOT synthetic
clusters.
We emphasize again that these results are not directly comparable
because, in the case of the \texttt{ASteCA} membership probabilities, we
also include the performance of the center of the cluster and radius estimation
methods.
If any of these fail, which is not uncommon for scarcely populated clusters or
those embedded in fields with large amounts of contamination, then the
Bayesian membership estimation method in \texttt{ASteCA} fails too.
This fact notwithstanding, this is another great result that demonstrates
the capabilities of pyUPMASK.

\end{appendix}

\end{document}